# Quantum-enhanced second harmonic generation beyond the photon pairs regime

Thomas Dickinson,[1,2,§] Ivi Afxenti,[3,§] Giedre Astrauskaite,[4] Lennart Hirsch,[3] Samuel Nerenberg,[4] Ottavia Jedrkiewicz,[2,5] Daniele Faccio,[4] Caroline Müllenbroich,[4] Alessandra Gatti,[5] Matteo Clerici,[2,3] Lucia Caspani[1,2,†]

[1] Institute of Photonics, Department of Physics, University of Strathclyde, Glasgow G1 1RD, United Kingdom

[2] Como Lake Institute of Photonics, Dipartimento di Scienza e Alta Tecnologia, Via Valleggio, 11, 22100 Como, Italy

[3] James Watt School of Engineering, University of Glasgow, Glasgow G12 8QQ, United Kingdom

[4] School of Physics and Astronomy, University of Glasgow, Glasgow G12 8QQ, United Kingdom

[5] Istituto di Fotonica e Nanotecnologie del CNR, Piazza Leonardo da Vinci 32, 20133 Milano, Italy

[§] These authors contributed equally, [†] lucia.caspani@uninsubria.it

**Abstract:** Two-photon processes are crucial in applications like microscopy and microfabrication, but their low cross-section requires intense illumination and limits, e.g., the penetration depth in nonlinear microscopy. Entangled states have been proposed to enhance the efficiency of two-photon interactions and have shown effectiveness at low intensities. This quantum enhancement is generally believed to be lost at high intensities, for more than one photon per mode, raising doubts about its usefulness. We explored experimentally and theoretically two-photon processes driven by entangled photons at intensities beyond this threshold and compared the results with the classical case. We found that a quantum advantage can still be observed at nearly one order of magnitude higher intensities than previously assumed. Our findings show a potential path for exploiting quantum-enhanced two-photon processes in practical applications.

**Introduction**

Two-photon interactions underpin several key applications, such as nonlinear imaging (*1*), deep tissue microscopy (*2*), spectroscopy, photodynamic therapy (*3*), data storage and microfabrication (*4*). In life science applications, such as deep tissue and functional imaging aimed at studying conditions like Alzheimer's disease and other nervous system disorders, increasing the penetration depth of the two-photon process is crucial for sampling in-vivo morphology and physiology deeper within tissues without causing damage (*2*). However, the low probability of two photons occupying the same interaction volume necessitates intense laser pulses, which can bleach fluorophores or damage the sample (*5*).

Entangled photon pairs have been suggested as a potential resource to overcome these limitations. They are a defining feature of Squeezed Vacuum (SV) states routinely produced by Parametric Down Conversion (PDC) in quadratic nonlinear media. Entangled photon pairs can enhance the probability of two-photon processes such as two-photon absorption (TPA) and second harmonic generation (SHG), transforming the interaction into a linear process with a large cross-section (*6–11*). Entangled TPA (eTPA) is expected to improve the sensitivity of imaging (*8*) and spectroscopy (*12–14*), where a reduction of the pump intensity can enable data acquisition with faint, less damaging illumination. It could also allow for the extraction of otherwise inaccessible information, e.g., on collective molecular resonances (*15*, *16*). Despite these promising predictions followed by experimental investigations (*17–20*), the advantages brought by eTPA have been debated in recent literature (*21–23*). Alternative (classical) mechanisms have been proposed to explain some of the observations (*24*, *25*). Most of the debate focused on the effective value of quantum enhancement and the lack of reproducibility of the experimental results (*22, 23, 26, 27*). These difficulties partially arise from the complexity of molecular fluorophores used in TPA experiments. Large fluorophores feature strong dependencies on temperature and concentration, along with complex excitation pathways and loss mechanisms, which are hard to control and model.

Up to now, the benefits of utilizing squeezed states to enhance two-photon processes have been demonstrated with spatially single-mode radiation and at low intensities, where the interaction volume contains less than a photon pair per event (*11, 28*). In this regime, the quantum advantage stems from the photon pairs behaving as a single particle, which results in a linear relation between the illumination power and the SHG/TPA signal. However, this necessitates low illumination powers and, in turn, results in low nonlinear signals that are of little practical use.

Here, we use sum-frequency processes to investigate two-photon interactions driven by broadband, largely multimode (spatially and temporally) squeezed radiation with a large range of photons per interaction volume. We demonstrate that quantum enhancement persists at illumination levels larger than previously thought (*29*), i.e., at more than one photon per interaction volume. Importantly, we quantify the quantum enhancement by direct comparison with the classical state typically used for TPA and SHG measurements. This is in contrast with most previous studies that compared the case of pure quantum illumination with that of quantum states spoiled by losses (*11, 28*) or conditioned by post selection to mimic classical radiation (*30*).

We focus on SHG, a simple and easily controllable process that, unlike TPA, does not involve intermediate states of excitation. Furthermore, SHG is a coherent process that can discriminate whether a two-photon signal is generated by entangled or nonentangled photons. This allows us to isolate the quantum enhanced signal (entangled-SHG, eSHG, a purely quantum process) from other contributions.

We first investigated the SHG process driven by multimode SV with a low average number of photons per mode (less than one), generated by low-gain PDC. By measuring the eSHG signal, we observed its expected linear scaling with the driving field intensity (*11*), which enabled us to establish the eSHG efficiency and analyze its dependence on optical losses applied to the SV. We then investigated the SHG driven by SV with many photons per mode, generated by high-gain PDC. It has been assumed that the deterministic recombination of entangled photons, which underpins the linear scaling mentioned above, would quickly become negligible in this regime (*11, 29*). We demonstrated that this is not the case. We further compared the efficiency of the SHG process driven by multimode SV to that of a classical coherent pulse (standard laser) with the same intensity (same energy, beam size, and pulse duration). We showed that SV drives the SHG process with higher efficiency than a coherent pulse, even when its modes are populated by more than one photon. This direct comparison sheds light on the levels of quantum enhancement that can be expected in real-case scenarios.

Our experimental findings are supported by a theoretical analysis that includes the full spatial and temporal properties of the PDC and the SHG process, predicting that the quantum enhancement is preserved for SV having more than one photon per mode (*31*). Considering our experimental parameters for the optical fields and the nonlinear crystals, the model achieves excellent agreement with the experiments, offering reliable guidance for future investigations.

**Results**

We have designed our experiment to investigate eSHG driven by pulsed, multimode SV with a controllable number of photons per mode (from below one to far above). The choice of a pulsed source differs from previous works (*11, 28*) and is required for the generation of SV with a high number of photons per mode (*23, 30*). A scheme of the setup is shown in Fig. 1A. A 515 nm pump generates a broadband SV via PDC in a 2 mm Beta Barium Borate (BBO) crystal, which is re-imaged on a second, identical crystal by an achromatic and dispersion-free telescopic system based solely on reflective optics (see *Methods – Generation of the quantum field*). The spatiotemporal entanglement volume, describing the region where an entangled photon pair is confined, is determined by the overall temporal and angular bandwidths (governed by phase-matching and/or spectral filters) and is roughly the inverse of the corresponding spectral volume. A pass-band filter (Brightline FF01-1055/70-25, Semrock) selects the temporal bandwidth of the SV ($\simeq$ 125 THz full-width at half maximum, FWHM, corresponding to $\simeq$ 70 nm around the central wavelength of 1030 nm – see Fig. 1B) to a region where the detrimental impact of the group delay dispersion (GDD) in the nonlinear media, optical elements and coatings is minimal. The angular bandwidth is limited by phase matching to $\simeq$ 300 mm$^{-1}$ (FWHM, corresponding to $\simeq$ 50 mrad, measured within a 10 nm bandwidth around 1030 nm), as shown in Fig. 1C. Therefore, the entanglement volume has a $\simeq$ 22 μm transverse spatial dimension and $\simeq$ 45 fs duration (FWHM). The second crystal up-converts the SV, and the generated signal around 515 nm is measured by a photomultiplier tube (PMT). The SHG process can also be driven by the combination of non-entangled photons (*32–34*), generating an incoherent second harmonic signal. In contrast, the eSHG process involving entangled photons results in a coherent signal with similar spatiotemporal coherence properties as the PDC pump (*35*). To select only this coherent signal, we filtered the SHG angular and temporal spectrum (see *Methods – Detection of the eSHG* for details).

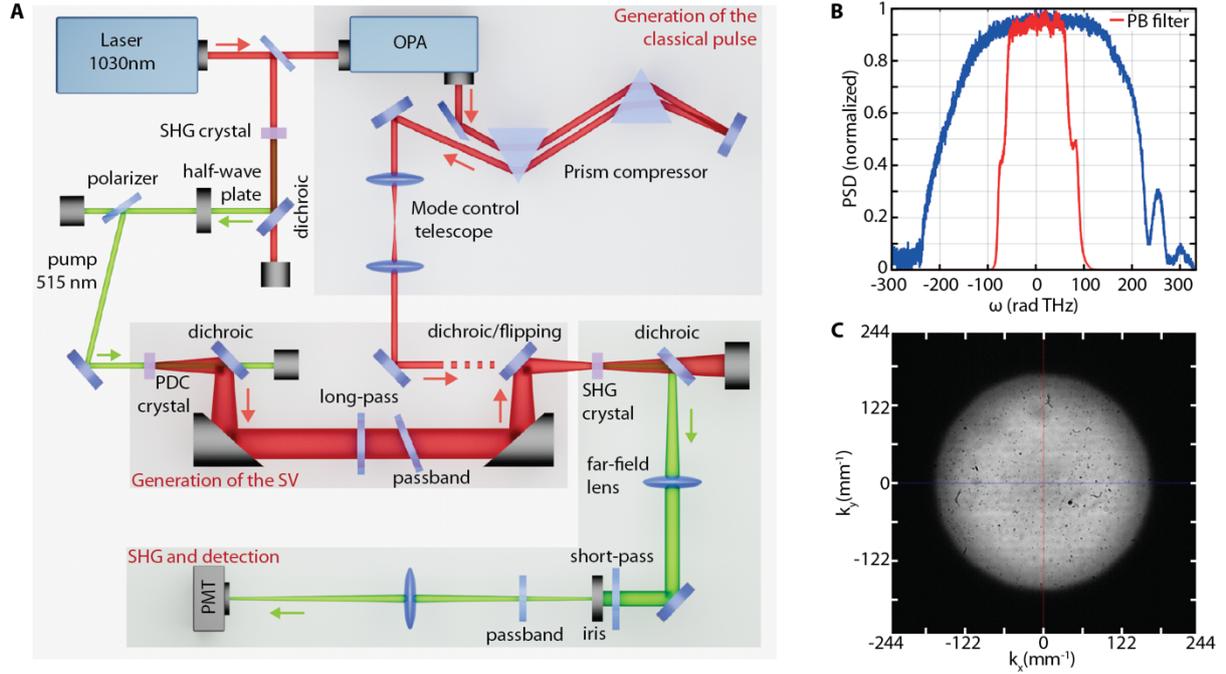

**Fig. 1 Experimental configuration. (A)** Sketch of the experimental setup with the three main building blocks responsible for the SV generation, second harmonic generation and detection, and the generation of the classical field for comparison. PMT: photomultiplier tube; OPA: optical parametric amplifier. **(B)** Power Spectral Density (PSD) of the SV transmitted by the optical system with (red) and without (blue) the passband (PB) filter. The latter is mainly constrained by the reflectivity of the dichroic mirrors employed to reject the 515 nm pump. **(C)** Angular spectrum of the SV after filtering with a 10 nm passband filter centered at 1030 nm.

**Characterization of the multimode SV.** As the first essential step towards comparing the entangled and classical SHG efficiency, we characterized the SV as a function of the PDC pump energy. This is the most direct control parameter for changing the SV number of photons per mode. To this end, we measured the average number of SV photons generated per pulse, $N_{SV}$, for increasing number of photons in the PDC pump pulse, $N_P$. The experimental data (Fig. 2A) clearly shows linear photon-pair generation at low PDC gains and nonlinear

(quadratic) photon-pair generation at higher PDC gains. We compared these results with the prediction of a quasi-stationary (QS) model for multimode pulsed PDC (*31, 36, 37*), using the hypotheses of negligible group velocity mismatch and spatial walk-off between the interacting fields. This model predicts that

$$N_{SV} = K_m(\Lambda\sqrt{N_P}) \sinh^2(\Lambda\sqrt{N_P}), \tag{Eq. 1}$$

where $\Lambda = 2\, l_c\, d_{\text{eff}} \sqrt{\frac{\hbar \omega_P \omega_{SV}^2}{2\varepsilon_0 n_P n_{SV}^2 c^3 V_P}}$, $l_c$ is the crystal length, $\omega_{P,SV}$ are the frequencies of the PDC pump and SV, $n_{P,SV}$ are their refractive indices in the nonlinear crystal, $d_{\text{eff}} \simeq 1.79$ pm/V (*38*) is the crystal second-order nonlinear coefficient, and $V_P$ is the pump spatiotemporal volume. We highlight the good matching between the theory and the experimental results, under the sole condition of reducing the effective nonlinearity to $d'_{\text{eff}} \simeq 1.65$ pm/V, i.e., a <10% difference from the tabulated value. Such a correction is compatible with the effect of the group velocity mismatch, not accounted for by the theoretical curve. In Eq. 1, $\Lambda\sqrt{N_P}$ is the dimensionless parametric gain, while $\sinh^2(\Lambda\sqrt{N_P})$ provides an estimate of the population of the modes (it would be the mean photon number in a simplistic model of single-mode SV at perfect phase-matching), and from now on will be referred to as the *number of photons per mode* $\langle n \rangle_m$. $K_m$ is a function of the gain (*31*) and can be roughly interpreted as the number of modes of the SV. In our case, it decreases at increasing gain (see inset in Fig. 2B) mainly due to the exponential shrinking of the SV profile (*36*). We characterized the SV profile experimentally, as shown in Fig. 2B for one spatial coordinate, obtaining a good match with the QS model predictions.

Intuitively, the number of SV modes is related to the ratio of the spatiotemporal size of the SV pulse to the entanglement volume. If we assume that the SV pulse profile matches that of the PDC pump, we get $K_m \simeq (185/45) \times (1500/22)^2 \simeq 19100$ modes. We note that $\langle n \rangle_m$ is often used to define the bound for the quantum-enhanced regime with the common assumption that the advantage is lost for $\langle n \rangle_m > 1$, and is commonly inferred from $N_{SV}$ assuming a constant number of modes. However, $K_m$ (calculated for our case and shown in the inset of Fig. 2B), while nearly constant in the low-gain PDC ($\Lambda\sqrt{N_P} \ll 1$), decreases steeply for high PDC gain. The value of $\langle n \rangle_m = N_{SV}/K_m$ shown in our plots is, therefore, obtained using the dependence of $K_m$ from the PDC gain given by the theory (*31*). In Fig. 2A we highlight with a green shaded area the region with $\langle n \rangle_m < 1$, where quantum enhancement has been mainly investigated so far. In this work, we show that the enhancement persists for $\langle n \rangle_m > 1$.

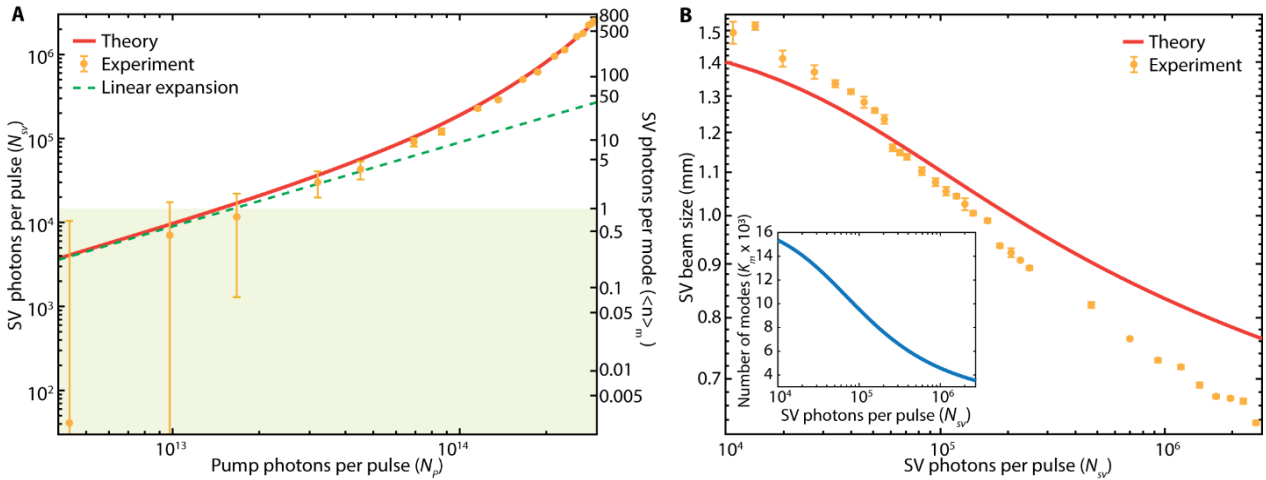

**Fig. 2. Squeezed Vacuum characterization.** (**A**) Log-log plot of the number of SV photons per pulse ($N_{SV}$) vs the number of pump photons per pulse ($N_P$). The red line is the expected trend predicted by our theoretical model (see text for details) while the green dashed line shows a linear approximation valid in the spontaneous regime. The green shaded area highlights the region where the average number of SV photons per mode $\langle n \rangle_m < 1$ (right axis). (**B**) Size of the SV beam (intensity FWHM) as a function of the number of SV photons per pulse ($N_{SV}$), showing a narrowing of the SV volume for increasing gain. The red curve shows the predicted values from the theoretical model. The inset shows the dependence of the number of modes ($K_m$) from the number of SV photons per pulse ($N_{SV}$) predicted by the model. We used this curve to evaluate $\langle n \rangle_m = N_{SV}/K_m(N_{SV})$.

**Loss analysis.** Following common practice, we initially investigated the effect of losses on the efficiency of eSHG driven by the SV. Losses degrade the entanglement underpinning quantum enhancement. As a result, a reduction in the eSHG efficiency for a degraded SV state, compared to an unspoiled one with the same mean number of photons per pulse at the SHG crystal, is considered a reliable indicator of quantum enhancement (*11*, *28*). We first focused on the $\langle n \rangle_m < 1$ regime and we compared the eSHG efficiency of an ideally pure SV with that of a modified state obtained by imposing different losses (30% and 50%) with neutral density filters.

The results for the unspoiled SV (Fig. 3A green) follow the expected linear trend. Each data point is an average over a 5-minute acquisition at 500 kHz, and the error bars show the standard deviation. A generation efficiency $\eta_{SV} = (3.3 \pm 0.2) \times 10^{-10}$ is obtained from the linear fit $N_{eSHG} = \frac{\eta_{SV}}{2} N_{SV}$ of our data, where $N_{eSHG}$ is the number of eSHG photons per pulse. This compares well with the prediction of the theoretical model $\eta_{SV,Th} = 4.3 \times 10^{-10}$ obtained considering only the linear contribution to eSHG (evaluated at $\langle n \rangle_m = 0.05$) and the reduced effective nonlinearity $d'_{eff}$. We note that the difference (~23%) can be ascribed also to residual losses, aberrations in the imaging system, and dispersion from materials and coatings decreasing the localization of entangled photon pairs in the SHG crystal.

The results obtained by introducing 30% and 50% losses are shown in blue and red, respectively. Note that, to maintain the same photon flux at the SHG crystal, the PDC gain was increased by increasing the pump pulse energy. In these two conditions, the SHG signal can still be fitted considering a linear dependence, albeit with lower generation efficiencies $\eta_{30\%} = (2.2 \pm 0.1) \times 10^{-10}$ and $\eta_{50\%} = (1.3 \pm 0.1) \times 10^{-10}$.

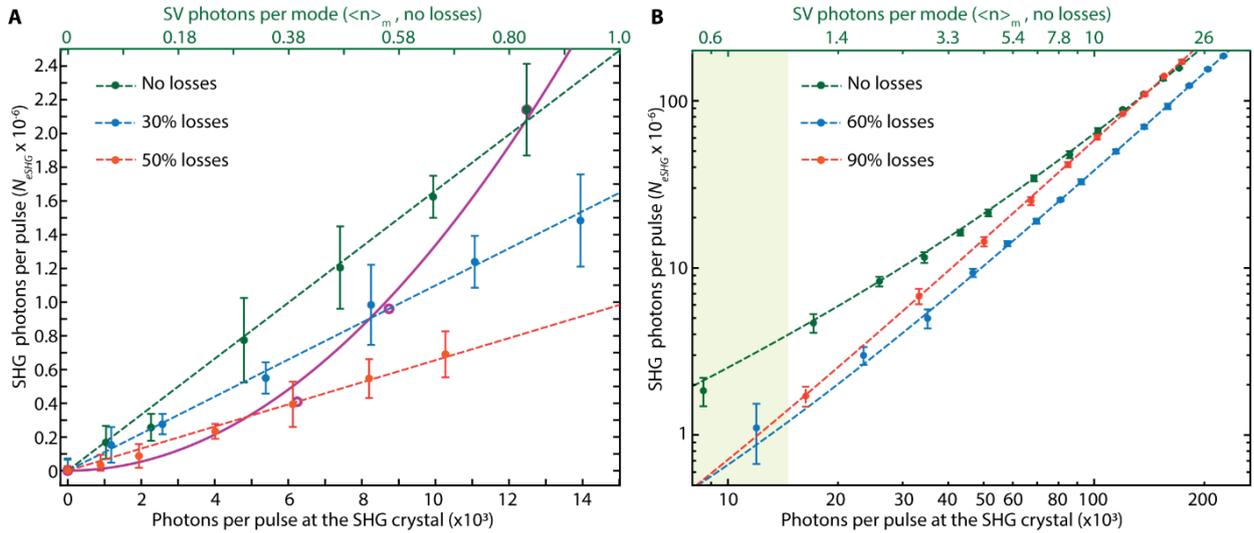

**Fig. 3. Loss analysis.** (**A**) Measurement of the SHG efficiency: number of SHG photons per pulse ($N_{SHG}$) vs number of the input photons on the SH crystal (bottom axis). The top axis reports the number of photons per mode, $\langle n \rangle_m$, for the unspoiled SV case (no losses). The green dots represent the SV input (no losses), while the blue and red dots are for 30% and 50% losses, respectively. The dashed lines are the best fits with a linear function. The purple circles identify the data extracted to have the scaling for fixed SV energy and increasing losses. They follow well a purely quadratic trend (solid purple curve, see text for details). (**B**) Same measurement as in (A) but for higher PDC gain plotted in log-log scale. In this case, we used higher losses (60% and 90% – blue and red, respectively) to enhance the effect. The dashed lines are best fit with a linear plus quadratic term ($y = ax + bx^2$). The green shaded area highlights the region where the unspoiled SV (no losses) has $\langle n \rangle_m \leq 1$.

To compare with the results available in the literature, we extracted the trend expected for the eSHG if the SV flux was decreased, for a fixed PDC gain, by introducing increasing losses. Specifically, we selected for the no-loss case the data point corresponding to the maximum number of photons per pulse at the SHG crystal ($\sim 12.6 \times 10^3$), for the 30% losses we selected the fit value at $0.7 \times 12.6 \times 10^3$, and for the 50% losses we selected the fit value at $0.5 \times 12.6 \times 10^3$. These are identified by purple circles in Fig. 3A and, as expected (*11*), match a purely quadratic trend (solid purple curve). This test is crucial to prove the quantum nature of the process (*24, 39*). It has indeed been shown in eTPA studies that the sole linear scaling for two-photon processes is not necessary evidence of a quantum effect as it could arise from residual single-photon events or by, e.g., hot-band absorption (*23–25*). Notably, from our eSHG results, it is also clear that a linear trend is still visible with spoiled quantum states.

We then increased the PDC pump energy to investigate the unexplored regime of eSHG driven by spatially and temporally multimode SV with $\langle n_m \rangle > 1$. The results for an unspoiled SV, i.e. with no additional losses, are shown in green in Fig. 3B together with those recorded by introducing 60% (blue) and 90% (red) losses. The dashed curves are fits obtained considering a linear plus a quadratic dependence of the SH signal from the input number of photons. Higher values of losses have been chosen here to showcase the competition between quantum correlations and the effects of the increased PDC gain.

We first notice that the efficiency for the no-loss SV is larger than that obtained with a spoiled SV well into the $\langle n \rangle_m > 1$ regime. However, it is also evident that such a trend does not continue indefinitely. Indeed, the eSHG efficiency for the 90%-loss case overcomes that from the no-loss SV at nearly $\langle n \rangle_m \simeq 15$. To understand this effect, it is necessary to recall that higher PDC gain is required to achieve the same photon flux at the SHG crystal for the spoiled SV. As shown before (Fig. 2B), this results in a narrowing of the generated (spoiled) SV field size. Such a reduction results in a higher peak intensity, which thus increases the SHG efficiency despite the reduced quantum correlations. The same effect is predicted by our theoretical model, see *Supplementary Materials – Theoretical analysis of the impact of losses*.

**Comparison with a classical field.** Following from our investigation on the role of losses, it is clear that to properly assess quantum enhancement, a comparison with a classical field is required. We therefore choose to compare the eSHG efficiency from SV with that of SHG produced by the classical state routinely employed for imaging, i.e., a standard laser pulse (a coherent state). The comparison was performed by matching the two intensity profiles, i.e., maintaining the same duration, beam size, and number of photons per pulse. Since the intensity profile of the SV changes with gain, we built a classical source with a tunable beam size and pulse duration, allowing matching to the SV properties, see *Methods – Generation of the classical field*.

The data for the SHG signal generated by the classical source vs the number of pump photons impinging on the crystal (Fig. 4A, red squares) matches well the theoretical model (*31*), shown with a solid, red curve. In the same figure, we show the results obtained using the SV (green circles) and the theoretical prediction (solid green curve). Both the classical (red) and the entangled (green) theoretical curves have been obtained after multiplying the bare theory results by the same coefficient 0.76, which amounts to the already mentioned correction for the $d_{\text{eff}}$ ($0.92^2$), and to a further 10% efficiency reduction that can be ascribed to losses and residual dispersion effects. It should be noted that in this comparison, the spectrum (spatial and temporal) of the classical pulse is much narrower than the SV one, giving all the advantages to the classical excitation. Yet, it is evident that the SV efficiency is still larger than that of a classical pulse well above the one-photon-per-mode regime, up to an average number of photons per mode $\langle n \rangle_m \simeq 9.3$, confirming our main claim.

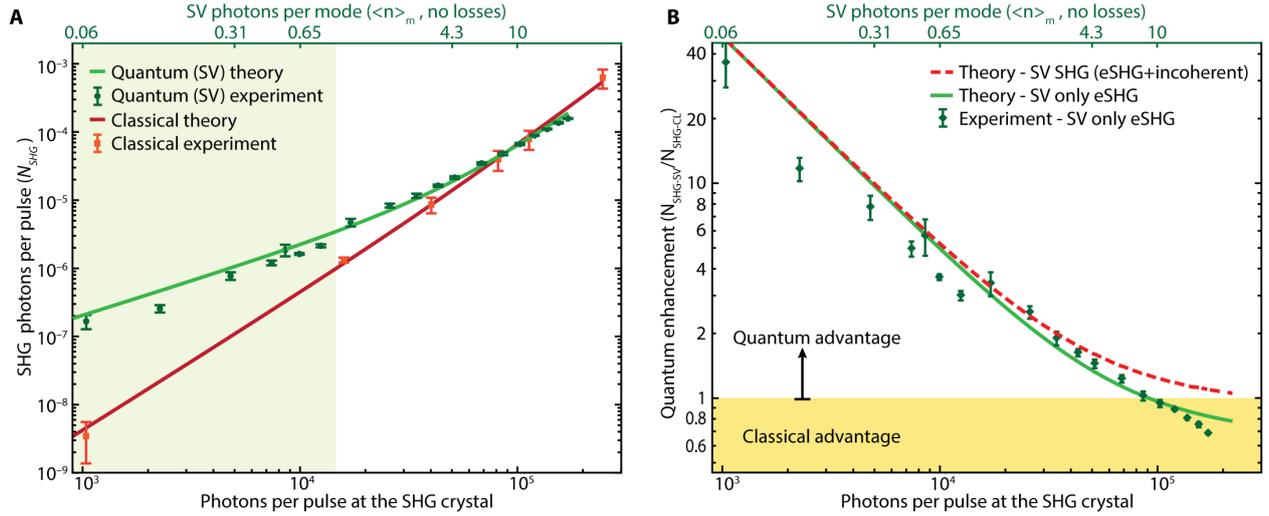

**Fig. 4. Comparison with a standard classical field (laser beam).** (A) SHG signal as a function of the number of photons per pulse at the SHG crystal for the quantum (green) and classical (red) case. A quantum advantage is maintained up to $\langle n \rangle_m \simeq 9.3$ (top axis). The solid curves are the theoretical predictions of the model (*31*) both obtained by considering a reduced $d'_{eff} \simeq 1.65$ pm/V and 10% overall losses on the SHG signal. The green-shaded area highlights the region where the SV has $\langle n \rangle_m \leq 1$. (B) Quantum enhancement, defined as the ratio between the SV-driven SHG and the SHG from a classical field (laser), i.e., $N_{SHG-SV}/N_{SHG-CL}$. The green diamonds show the ratio between the experimental eSHG values (green circles in **A**) and the theoretical prediction for classical SHG (red curve in **A**). The solid green curve is the theoretical prediction (ratio between green and red curves in **A**). The dashed red curve is the theoretical prediction for the total enhancement the SV brings to SHG (coherent and incoherent components, see text for details). The yellow-shaded area illustrates the region where the classical case outperforms the SV.

It should also be noticed that the experimental data were obtained by carefully optimizing, at each different PDC gain, the position of the SHG crystal with respect to the image plane of the PDC crystal, so to maximize the eSHG rate (the same optimization was done in the theory). The quantum enhancement obtained using the SV is shown in Fig. 4B, calculated as the ratio between the experimental values of the eSHG and the theoretical values for the classical SHG. In the same graph, we also show the theoretical prediction (solid green curve).

**Discussion**

The results in Fig. 4A do not include the full SHG signal generated by the SV. Instead, they show the coherent signal induced solely by the up-conversion of entangled photons, namely the eSHG. As mentioned above, the incoherent (coherent) component is spread over a large (narrow) wavevector spectrum so that the selection is experimentally performed by inserting an iris in the far field of the SH signal and filtering a narrow portion of the SH spectrum as shown in Fig. 1A. The eSHG signal has a linear component that dominates for low number of SV photons per mode and grows quadratically at increasing SV intensities, leading to an overall quantum enhancement up to $\langle n_m \rangle \simeq 9.3$. Remarkably, the experimental results match well the outcomes of the theoretical model (*31*) for both the classical and the quantum cases. It should be noted that the theory predicts that the incoherent SH arising from the up-conversion of non-entangled SV photons adds an additional classical quadratic term (*30*). When the contribution from non-entangled photons is also included, the SHG from SV have a larger efficiency than that of a classical field with the same profile independently from the SV intensity (*31*), as shown by the red dashed curve in Fig. 4B.

Our study demonstrates a significant gain arising from quantum effects in two-photon processes at intensities one order of magnitude larger than what was previously considered feasible. Furthermore, the good agreement between the theory and the experimental results supports the choice of SV as a more efficient field to drive two-photon processes even at intensities for which any quantum advantage was expected to become negligible (*11*, *35*).

**Materials and Methods**

**Generation of the quantum field.** The pump pulse for the PDC had a 1.5 mm diameter beam size (FWHM of the intensity profile), a $\simeq$ 185 fs (FWHM) pulse duration, $\simeq$ 515 nm central wavelength, and a variable 200/500 kHz repetition rate, and was obtained by frequency doubling in a 1-mm-thick Beta Barium Borate crystal (BBO, Light Conversion) the $\simeq$ 245 fs duration, 1030 nm central wavelength output of a Yb:KGW amplified laser (Carbide CB3 40W, Light Conversion). The energy of the PDC pump could be tuned by rotating an achromatic half-wave plate in front of a thin film polarizer (Eksma) in the 1–120 µJ/pulse range. The pump was then injected into a 2-mm-long BBO crystal cut for collinear Type I SHG of a 1030 nm field coated with anti-reflection layers (Eksma), where it drove the PDC process resulting into the generation of a broadband multimode SV. This radiation was re-imaged into a second BBO crystal identical to the one used for PDC using a confocal imaging system consisting of two 8-inch equivalent focal length, 1-inch aperture, 90 deg off-axis parabolic silver mirrors with broadband ultrafast coatings (low GDD, Edmund Optics) arranged in a $4f$ configuration. The beam path was folded employing 7 low dispersion dichroic mirrors (Layertec, not all shown in the figure) with high reflectivity (99.9% at an incidence angle of 45 deg) in a 300 nm bandwidth centered at 1030 nm and with high transmission (>99%) at 515 nm. The low dispersion is required to maintain the temporal localization of the entanglement volume of the SV radiation. To further remove the intense PDC pump (at 515 nm) from the SV, a long-pass filter (FELH0850, Thorlabs) with high optical density at 515 nm was inserted in the beam path. Finally, a high transmissivity band-pass filter (Brightline FF01-1055/70-25, Semrock) was also inserted and tuned at $\simeq$ 23 deg to cut the SV spectrum symmetrically around the central wavelength of 1030 nm (and to increase the pump rejection further). This way, a sufficient rejection was ensured to avoid contaminating the eSHG signal with residual PDC pump photons. The SV number of photons per pulse $N_{SV}$ was measured with a high-sensitivity calibrated photodiode (S132C, Thorlabs) and background-corrected to remove the contribution from parasitic radiation, e.g., scattering from the laser source.

**Detection of the eSHG.** After the SHG crystal, the residual SV was removed from the beam path with four dichroic mirrors of high reflectivity at 515 nm and high transmission over a large bandwidth (>80 nm) around 1030 nm (Eksma), so that the up-converted signal could be measured with a photomultiplier tube (PMT, H16722P-40, Hamamatsu, 40% quantum efficiency at 515 nm – data rescaled accordingly) connected to a time-to-digital converter (HydraHarp 400, PicoQuant) for the time-gated signal acquisition synchronized to the source pulse train. To select only the coherent signal (eSHG), we filtered the SHG signal by inserting a 1 mm diameter aperture into the radiation far-field (generated with an $f = 200$ mm focal length lens confocal to the eSHG crystal) and, analogously, by employing a narrow-band pass-band filter (10 nm bandwidth around 515 nm). We note that the incoherent SHG radiation in the far-field extended over a ~12-mm-diameter area, therefore its contribution to the recorded signal (i.e., captured by the 1 mm aperture) was negligible (<0.7%). The SH counts were corrected by subtracting a background measured by rotating the SHG crystal by $\pi/2$ around the surface normal, i.e., in a condition where no conversion occurs.

**Generation of the classical field.** To generate a coherent classical pulse with the same intensity profile and energy of the SV at different energies, we employed the temporally compressed output from an optical parametric amplifier (OPA, Orpheus-F, Light Conversion), delivering 1030 nm pulses of $\simeq$ 120 fs duration. The duration was changed with a prism compressor to match the expected duration of the SV at different energies. Similarly, the beam size was adjusted with telescopes to match the measured SV beam width. In Table 1 we report the values for the energy, pulse duration (measured with a home-built autocorrelator), and beam size of the classical pulse used in the comparison.

Table 1. Parameters of the classical field.

| Photons at the SHG crystal ($10^3$/pulse) | Beam size (mm, FWHM) | Pulse duration (fs, FWHM) |
|---|---|---|
| 1.0 | 1.48 $\pm$ 0.05 | 173 $\pm$ 5 |
| 15.9 | 1.33 $\pm$ 0.05 | 170 $\pm$ 5 |
| 40.2 | 1.16 $\pm$ 0.05 | 165 $\pm$ 5 |
| 81.6 | 1.07 $\pm$ 0.05 | 158 $\pm$ 5 |
| 113.1 | 1.03 $\pm$ 0.05 | 153 $\pm$ 5 |
| 246.9 | 0.90 $\pm$ 0.05 | 137 $\pm$ 5 |

**Acknowledgments**

Innovate UK, BQS, project reference 10075401 (DF, MC)
UKRI, Fellowship "In-Tempo", EP/S001573/1 (MC)
Royal Academy of Engineering, Chair in Emerging Technologies (DF)
EPSRC, EP/Y029097/1 (DF)
EPSRC, EP/V051148/1 (CM)
European Commission (ERC) Consolidator Grant, QuNIm, G.A. 101125923 (MC, LC)
UKRI and EPSRC, EP/V062492/1 (LC).
UKRI and EPSRC, "QuantIC", EP/T00097X/1 (IA, CM, MC, LC)
Fraunhofer UK PhD funding (TD).
PRIN 2022K3KMX7 of the MUR ELISE, CUP B53D2300515000 (TD, OJ, AG, MC)

**Author contributions:**
Conceptualization: DF, CM, AG, MC, LC
Methodology: TD, IA, OJ, DF, CM, AG, MC, LC
Formal analysis: TD, IA, GA, LH, AG, MC, LC
Investigation: TD, IA, GA, SN, MC, LC
Theoretical model: AG
Writing—original draft: TD, IA, AG, MC, LC
Writing—review & editing: All the authors
Visualization: TD, IA, AG, MC, LC
Supervision: DF, CM, MC, LC
Project administration: MC, LC